\documentclass[aps,prd,twocolumn,preprintnumbers,superscriptaddress,floatfix]{revtex4}

\usepackage{multirow, graphicx,amssymb,url,mathrsfs,amsmath}
\usepackage{eucal,wrapfig,boxedminipage,setspace,subfigure}
\usepackage{amsxtra,amstext,latexsym,dsfont}

\def\a  {\alpha}       \def\b  {\beta}

 \newcommand{\call}{\mbox{${\cal L}$}}

\def\IR{{\hbox{{\rm I}\kern-.2em\hbox{\rm R}}}}
\def\IB{{\hbox{{\rm I}\kern-.2em\hbox{\rm B}}}}
\def\IN{{\hbox{{\rm I}\kern-.2em\hbox{\rm N}}}}
\def\IC{\,\,{\hbox{{\rm I}\kern-.59em\hbox{\bf C}}}}
\def\IZ{{\hbox{{\rm Z}\kern-.4em\hbox{\rm Z}}}}
\def\IP{{\hbox{{\rm I}\kern-.2em\hbox{\rm P}}}}
\def\IH{{\hbox{{\rm I}\kern-.4em\hbox{\rm H}}}}
\def\ID{{\hbox{{\rm I}\kern-.2em\hbox{\rm D}}}}

\def\half{\frac{1}{2}}

\def\Tr{{\rm Tr}\,}

\newcommand{\eg}{{\it e.g.}}
\newcommand{\ie}{{\it i.e.}}

\newcommand{\beq}{\begin{equation}}
\newcommand{\eeq}{\end{equation}}
\newcommand{\bea}{\begin{eqnarray}}
\newcommand{\eea}{\end{eqnarray}}

\newcommand{\tz}{t^{(0)}}
\newcommand{\MP}{M_{\rm Planck}}
\newcommand{\EV}{E_{\rm vac}}
\newcommand{\Es}{\Lambda_{\rm stripe}}

\begin{document}

\voffset 1cm

\newcommand\sect[1]{\emph{#1}---}

\title{Translational Symmetry Breaking in Higgs \& Gauge Theory, \\and the Cosmological Constant}

\author{Nick Evans}
\email{evans@soton.ac.uk}
\affiliation{STAG Research Centre \&  Physics and Astronomy, University of
Southampton, Southampton, SO17 1BJ, UK}
\author{Tim R. Morris}
\email{T.R.Morris@soton.ac.uk}
\affiliation{STAG Research Centre \&  Physics and Astronomy, University of
Southampton, Southampton, SO17 1BJ, UK}
\author{Marc Scott}
\email{M.Scott@soton.ac.uk}
\affiliation{STAG Research Centre \&  Physics and Astronomy, University of
Southampton, Southampton, SO17 1BJ, UK}

\begin{abstract}
We argue, at a very basic effective field theory level, that higher dimension operators in scalar theories that break symmetries at scales close to their ultraviolet completion cutoff, include terms that favour the breaking of translation (Lorentz) invariance, potentially resulting in striped, chequered board or general crystal-like phases. Such descriptions can be thought of as the effective low energy description of QCD-like gauge theories near their strong coupling scale where terms involving higher dimension operators are generated. Our low energy theory consists of scalar fields describing operators such as $\bar{q} q$ and $\bar{q} F^{(2n)} q$. Such scalars can have kinetic mixing terms that generate effective momentum dependent contributions to the mass matrix. We show that these can destabilize the translationally invariant vacuum.  It is possible that in some real gauge theory such operators could become sufficiently dominant to realize such phases and it would be interesting to look for them in lattice simulations. We present a holographic model of the same phenomena which includes RG running.  A key phenomenological motive to look at such states is recent work that shows that the non-linear response in $R^2$ gravity to such short range fluctuations can mimic a cosmological constant. Intriguingly in a cosmology with such a Starobinsky inflation term, to generate the observed value of the present day acceleration 
would require stripes at the electroweak scale. Unfortunately, low energy phenomenological constraints on Lorentz violation in the electron-photon system appear to strongly rule out any such possibility outside of a disconnected dark sector. 
\end{abstract}

\maketitle

\newpage

\section{Introduction}

Translational invariance is known to be spontaneously broken in a number of superconducting cuprate systems \cite{cuprates}. They display phases where the condensate varies spatially as $\sin k x$ manifesting as visible stripes in some measurements. The existence of such translationally non-invariant phases have also been speculated to exist in finite density gauge theory \cite{fdqcd1, fdqcd2, fdqcd3}.  There has been some work recently on modelling such phases in holographic descriptions of superconductors and finite density QCD \cite{Nakamura:2009tf,Ooguri:2010xs,Bayona:2011ab,Bergman:2011rf,Donos:2011qt}.  Two dimensional chequer board patterns are also possible \cite{Withers:2014sja}.The chemical potential in these systems already breaks Lorentz invariance and provides a natural Lorentz frame for stripes to form. Here though we wish to ask whether spontaneous breaking of Lorentz invariance, in this pattern, can occur in scalar or gauge theories at zero chemical potential (see \cite{Kostelecky:1988zi} for a well known related discussion of Lorentz violation in string theory). 

A preference for spatially dependent vevs (vacuum expectation values) for operators essentially requires that the relevant operators have negative kinetic terms in the unbroken vacuum which manifest in the effective potential as a negative $k^2$ dependent contribution to the mass term. Normally this is associated with ghost like behaviour and seems forbidden at weak coupling. We will argue though that it can happen in a theory where many higher dimension operators are present and are sufficiently large that when symmetry breaking occurs they generate effective terms that mimic negative kinetic terms. The true vacuum will then be characterized by Lorentz breaking vevs and fluctuations will then be ghost free in the true vacuum. Again in principle higher dimension operators evaluated in the striped vacuum can correct the signs and leave a stable theory. One could therefore imagine a Higgs-like theory with condensation occuring close to its UV (ultraviolet) cutoff scale displaying dynamical Lorentz invariance breaking. A natural environment for such an effective theory is the strong coupling regime of a gauge theory. At the scale of strong coupling many higher dimension operators 
become important and simultaneously chiral condensation and condensation of $\Tr F^2$ occurs. It at least seems possible that within the space of gauge theories, Lorentz symmetry breaking dynamics might exist. Our effective field theory discussions will not prove that any particular theory will behave in this way but it is a novel possibility that should be borne in mind in lattice simulations of models beyond the Standard Model. Alternatively, in gauge theories without translational symmetry breaking, one can reinterpret our results as bounds on the sizes of certain higher dimension operators in the effective theory. 

We will also present a holographic model (in the spirit of \cite{Erlich:2005qh, DaRold:2005zs} and more recently
\cite{Jarvinen:2011qe,Alho:2013dka}) of the instability. We describe the operators $\Tr F^2$, $\bar{q}q$ and  $\bar{q} F^2 q$ and represent their running anomalous dimensions as running mass squareds for the appropriate scalars in AdS space. The UV of the theory is stable and has zero operator vevs. As the Brietenlohner Freedman (BF) bound \cite{Breitenlohner:1982jf} is violated in the IR (infrared), condensation occurs and, if suitable ($k^2$ dependent) potentials are chosen, an instability for Lorentz violating vevs can emerge.  

Whilst the possibility of Lorentz violation is intriguing in itself,  we also present a more explicit phenomenological motivation. It has recently been shown \cite{Preston:2014tua} that in $R^2$ gravity short distance fluctuations in the metric can be converted by the non-linearities of the Einstein's equations into an effective long distance cosmological constant. Our interest in Lorentz violating vacua is partially motivated by thinking about how to generate such short distance fluctuations with sufficient power. Intriguingly if one considers this mechanism in Starobinsky early-universe inflation models where the $R^2$ term is set by the scalaron scale of $M\sim10^{13}$ GeV, then stripes at the electroweak scale generate the observed cosmological constant! 

Could gauge theories close to the Standard Model involve Lorentz violation then? The answer is fairly strongly no.  Limits on Lorentz violation \cite{Pospelov:2004fj,Pruttivarasin:2014pja} in the electron photon system are extremely stringent and constrain any coupling of such a system to be associated with very high scales. Therefore, if stripes are the source of the observed cosmological constant then they must be well hidden in a dark sector. 

One might also presume that the spontaneous breaking of Lorentz symmetry would generate Goldstone poles in the non-relativistic propagators of the theory - clearly no such massless modes exist in the visible Universe. In fact the number of long range propagating Goldstone modes depends on the pattern of symmetry breaking as has been discussed in \cite{Low:2001bw}. There are massless modes associated with each broken direction of translation but they only propagate along unbroken directions transverse to the breaking  (this may be familiar to the reader in the case of D-branes where massless modes result from the dimensional reduction of the 10d gauge field from the open string sector: they are tied to the D-brane's world-volume). Thus there will be long-range propagating Goldstones for striped or chequer board configurations where there still exist unbroken directions but not for cuboid or general crystal-like configurations (where the second-derivative of the potential vanishes only at a lattice of points and hence resembles an array of D0 branes). We will not exhibit these Goldstone structures here since we concentrate on the instability for the formation of stripes rather than a full model of the final ground state (an example of such massless phonons in the $\bar{q}q$ scalar is computed recently at high density in \cite{Hidaka:2015xza}). Since phenomenologically the Lorentz breaking sector must be extremely weakly coupled to the visible sector the presence of Goldstones in anycase is probably not an issue. 

Finally we note that we have considered whether striped ground states are ruled out in QCD-like gauge theories by the theorems of Vafa and Witten \cite{Vafa:1983tf,Vafa:1984xg}. For example, one theorem \cite{Vafa:1983tf} asserts that any state associated with the $\bar{q} q$ operator must be heavier than the pion; given the pion may be made massive by a small explicit quark mass, breaking of vector symmetries is, for example, forbidden. This may indeed forbid the appearance of striped and chequer board phases in vector like gauge theories where there will be Goldstone modes able to propagate in some directions but does not clearly prevent cuboid phases where the Goldstones can not propagate. (Chiral gauge theories which have a sign problem are not constrained by the theorem so for example ``moose'' \cite{Georgi:1985hf} with two QCD-like gauge theories connected by chiral fermions transforming under each group are examples of escapes from these theorems although they are untestable on the lattice). A second theorem  \cite{Vafa:1984xg} proves that parity cannot be spontaneously broken in vectorlike theories. Depending on the form of the translational symmetry breaking and the placement of the origin, parity is strictly speaking broken in these cases. However the theorem does not apply to these cases because it explicitly assumes that the operators sensitive to such parity breaking are Lorentz invariant operators $\int\!\! d^4x X(x)$ which are integrated over all space-time (and thus involve the totally antisymmetric tensor $\epsilon^{\mu\nu\alpha\beta}$). In fact once the operators are integrated over all space, the parity-breaking is no longer visible, since the modulated phase can effectively be translated and rotated by a change of integration coordinate. (The theorem therefore shows that such modulated phases must be invariant under parity compensated by translations and rotations, or more simply --assuming that charge conjugation and CPT are conserved-- that $T$ invariance is not broken.)


\section{Effective Higgs Theories}

Let us begin by writing down the simplest possible Higgs theory with one scalar and to quadratic order to demonstrate the usual instability
\begin{equation} 
{\cal L} = \partial^\mu \phi^* \partial_\mu \phi - V(|\phi|), \hspace{0.5cm} V = -m^2 |\phi|^2 \,.
\end{equation}
Now if we consider a ground state where the vev of the scalar has a striped vev in one direction ($\phi = v \sin kx$) then there is an effective potential
\begin{equation}
V = -(m^2-k^2) |\phi|^2 \,.
\end{equation}
Non-zero $k$ reduces the strength of the potential instability and is disfavoured. We can see that for there to be an 
instability that favours stripes we need to reverse the sign of the scalar kinetic term. However, we can not simply 
flip the sign on the kinetic term since the theory would become ill behaved with ghosts.

A scalar theory though is known to suffer from a hierarchy problem and the na\"\i ve expectation is that new physics will enter at a scale reasonably close to the scalar's mass; we will call this somewhat higher scale $\Lambda_{UV}$. The expectation is that at the scale $\Lambda_{UV}$ higher dimension operators will generically be present. Such higher dimension operators can, once symmetry breaking is triggered, lead to effective kinetic terms that favour translational symmetry breaking. For example, let us consider including an additional scalar $f$. We can imagine a term
\begin{equation} 
\Delta {\cal L} = - {\kappa_0 \over \Lambda_{UV}^2} |f|^2\partial^\mu \phi^* \partial_\mu \phi\,.
\end{equation}
Were $f$ to condense at some scale and $\kappa_0$ be large enough then the reversal of the kinetic term's sign is achieved. 

Once a striped phase has condensed other higher dimension operators can step in to secure the ghost inducing negative $(\partial_t \phi)^2$ term is not present in the true vacuum. For example, consider the term
\begin{equation} \Delta {\cal L} = {\kappa_1 \over \Lambda_{UV}^4} |\partial^\mu \phi|^2  |\partial^\nu \phi|^2  \end{equation} 
evaluated on the symmetry breaking solution (a Lorentz invariant term results if the vev occurs twice in one derivative term but a spatially prefering term occurs if the two vevs occur in the different derivative terms). This term will distinguish the spatial directions in which there are stripes from the temporal direction and the coefficient could be concocted to cure the ghost problem once the stripy vev had formed.

Of course, in this discussion many other terms might be present that oppose the effect, or indeed $\kappa_0$ might be small or negative. We simply wish to identify terms that could trigger translational symmetry breaking. 
Another possible mechanism is to introduce yet another new scalar, $\chi$, with the same symmetry properties as the original $\phi$. 
Now consider the terms
\begin{equation} 
\Delta {\cal L} = |\partial^\mu \phi|^2 +  |\partial^\mu \chi|^2 + m^2 |\phi|^2 - M^2|\chi|^2 
+ {\kappa_2 \over \Lambda_{UV}^2} |f|^2 \partial^\mu \phi^* \partial_\mu \chi
\end{equation}
Were $f$ to get a vev then an off-diagonal kinetic mixing is induced for the $\phi, \chi$ pair. The effective $k$ dependent quadratic potential is then given by
\begin{equation} 
(\phi, \chi) \left( \begin{array}{ccc} -m^2 + k^2 &   {\kappa_2 \over \Lambda_{UV}^2} \langle f \rangle^2 k^2\\
 {\kappa_2 \over \Lambda_{UV}^2} \langle f \rangle^2 k^2 & M^2 + k^2 \end{array} \right) \left( \begin{array}{c} \phi \\ \chi \end{array} \right) \,.\end{equation} 
For small $k$ the negative mass squared eigenvalue becomes
\begin{equation} m_1^2 = -m^2 + k^2 - {1 \over 2} {\left( {\kappa_2 \over \Lambda_{UV}^2} \langle f \rangle^2 k^2
\right)^2  \over M^2 -m^2} \,.
\end{equation} 
Again for not unreasonable choices of parameters this term could be made to favour translational symmetry breaking. Of course this is an argument for an instability rather than a full model of the final vacuum. The potential at large $k^2$ would need to be stabilized by terms with higher powers and the dynamically determined  value of $k$ may lie close to $\Lambda_{UV}$. The precise form of the vacuum is also dynamically determined - one could envisage 1d stripes, 2d chequer board patterns or 3d cuboid patterns. 

Such Lorentz violation would have to dynamically pick a frame of reference in our Universe, however, it seems likely that the innate frame of the matter in the Universe that now gives the frame of the 3K cosmic microwave background radiation would be chosen. As the gauge theory cooled and condensed the small chemical potential of the Universe would be the only parameter biasing a specific frame.

Such scalar models with $\phi, \chi$ and $f$ may look baroque. To argue that this is a sensible arena for discussion we should recast this analysis as the effective description of a QCD-like gauge theory. Consider an SU($N_c$) gauge theory and consider a single quenched quark in that theory when the number of flavours $N_f \ll N_c$. We know that the vacuum has a non-zero value of the quark condensate $\bar{q} q$ which carries U(1)$_A$ charge of 2 (we neglect the anomaly here). This operator should be mapped to $\phi$. We also know that the operator $\Tr F^2$ is non-zero in the vacuum and a singlet under flavour symmetries. It is the scalar $f$ above. Finally $\chi$ could represent the higher dimension operator of the form $\bar{q} F^2 q$  (or possibly those with higher powers of $F$): this operator has the same symmetry properties as $\bar{q} q$ but in the quantum theory is a distinct operator whose vev should be determined by the effective theory. In fact above we assumed that the $\chi$ field does not condense but simply mixes with $\phi$. 

In such asymptotically free theories the running coupling enters a regime of strong coupling at some scale which should be associated with the cutoff $\Lambda_{UV}$ of the scalar theory. At this scale the strong coupling is expected to generate higher dimension operators including of the form we have discussed above. The chiral condensate will then form in QCD quite quickly in RG running.

These arguments map the dynamics of strongly coupled gauge theories to the scalar models discussed above and suggest that translational symmetry breaking is at least possible in the vacuum. Of course we have in no way proved the phenomena occurs or is even likely. However, given the wide range of asymptotically free gauge theories that can be constructed it is possible that amongst them are some that do concoct their higher dimension operator couplings to conspire to this end. It would certainly be interesting to find such a theory on the lattice. 

In the next section we will construct a holographic model of a gauge theory's dynamics that reproduces this line of argument  and more carefully takes into account the scaling dimensions and RG flow in such a theory.

\section{A Holographic Model} 

To demonstrate the effective field theory arguments above a little more robustly in this section we will construct an AdS/QCD style holographic model \cite{Erlich:2005qh,DaRold:2005zs,Jarvinen:2011qe,Alho:2013dka}. It will show the possible translational symmetry breaking instability of a QCD-like gauge theory we discussed above. We assume there is some SU($N_c$) gauge theory with a small number of quenched quarks. As usual we place the effective theory in AdS$_5$
\begin{equation}
ds^2 = {dr^2 \over r^2} + r^2 dx_{3+1}^2 \,.
\end{equation}
We assume the underlying Yang Mills theory generates a vev for the operator $\Tr F^2$ and represent that by a background field in an AdS$_5$ space 
\begin{equation} f = {c \over r^4}\,. \end{equation}
Our model will concentrate on the quenched quark sector rather than the generation of this vev.
Although we will allow the AdS space to extend to $r= \infty$ such a gravity description should really only extend to the UV cutoff where the asymptotically free theory enters strong coupling. Experience teaches us that the models still work well without a UV cutoff because the dynamics is determined around the scale of the BF bound violation. For example we might expect $c \simeq (0.1-1 \Lambda_{UV})^4$. 

We now move to the study of the behaviour of the $\bar{q} q$ operator of the theory. Our model is based on the Dynamic AdS/QCD model of \cite{Alho:2013dka}. We represent $\bar{q} q$ by a field $X$ with action
\begin{equation}
S   =   \int d^4x~ d \rho\,  \rho^3 
\left[  {1 \over r^2} |D X|^2  
 +  {\Delta m^2 \over \rho^2} |X|^2    \right], 
\label{daq}
\end{equation} 
$r^2=\rho^2 + |X|^2$.
If $\Delta m^2 =0$ then the scalar, $X$, describes a dimension 3 operator and dimension 1 source as is required for it to represent $\bar{q} q$ and the quark mass $m$. That is, in the UV the solution for the $X$ equation of motion is $|X| \sim m + \bar{q}q/\rho^2$. We will work in the chiral limit with the quark mass zero henceforth. A non-zero $\Delta m^2$ allows us to introduce an anomalous dimension for this operator, $\gamma$. If the mass squared of the scalar violates the BF bound of -4 ($\Delta m^2=-1$, $\gamma=1$) then  the scalar field $X$ becomes unstable and the theory enters a chiral symmetry breaking phase. We will fix the form of $\Delta m^2$ using the two loop perturbative running of the gauge coupling in QCD with $N_f$ flavours transforming under a representation $R$. Of course this is a crude approximation to the running of the anomalous dimension $\gamma$ but it serves as a reasonable guess. This takes the form
\begin{equation} 
\mu { d \alpha \over d \mu} = - b_0 \alpha^2 - b_1 \alpha^3,
\end{equation}
where
\begin{equation} b_0 = {1 \over 6 \pi} \left(11 C_2(G) - 4N_fC_2(R)\frac{\text{dim}(R)}{\text{dim}(G)}\right), \end{equation}
and
\begin{equation} \begin{array}{ccl}b_1 &= &{1 \over 8 \pi^2} \left(\frac{34}{3}\left[C_2(G)\right]^2 \right.\\
&&\\ && \left.-\left[\frac{20}{3}C_2(G)C_2(R)+4\left[C_2(R)\right]^2\right]N_f\frac{\text{dim}(R)}{\text{dim}(G)}\right) \end{array} . \end{equation}
The one loop result for the anomalous dimension of the quark mass is
\begin{equation} \gamma_1(\mu;R) = {3 C_2(R) \over 2\pi}\alpha(\mu;R).  \end{equation}

We will identify the RG scale $\mu$ with the AdS radial parameter $r$ in our model. Note it is important that $X$ enters here. If it did not and the scalar mass was only a function of $\rho$ then, were the mass to violate the BF bound at some $\rho$, it would leave the theory unstable however large $X$ grew. Including $X$ means that the creation of a non-zero but finite $X$ can remove the BF bound violation leading to a stable solution.

Working perturbatively from the AdS result $m^2 = \Delta(\Delta-4)$ we have
\begin{equation} \label{dmsq3} \Delta m^2 = - 2 \gamma_1(\mu;R) = -{3 C_2(R) \over \pi}\alpha(\mu;R).\end{equation}
This will then fix the $r$ dependence of the scalar mass through $\Delta m^2$ as a function of $N_c$ and 
$N_f$ for each $R$. 

The Euler-Lagrange equation for the  vacuum embedding $X$ is given at fixed $\Delta m^2$ by the solution of 
\begin{equation}\label{embedeqn}
 \frac{\partial}{\partial\rho}\left( \rho^3 \partial_\rho X\right)  - \rho \Delta m^2 X =0.
\end{equation}
Note that if $\Delta m^2$ depends on $X$ at the level of the Lagrangian then there would be an additional term  $- \rho X^2 \partial \Delta m^2 / \partial X$. We neglect this term and instead impose the running of $\Delta m^2$ at the level of the equation of motion. The reason is that the extra term introduces an effective contribution to the running of $\gamma$ that depends on the gradient of the running coupling. Such a term is not present in perturbation theory in our QCD-like theories; we wish to keep the running of $\gamma$ in the holographic theory as close to the perturbative guidance from the gauge theory as possible.

In order to find $X(\rho)$ we solve the equation of motion numerically with shooting techniques with an input IR initial condition. A sensible first guess for the IR  boundary condition is
\begin{equation}\label{bca}  X(\rho=X_0) = X_0, \hspace{1cm}  X'(\rho=X_0)=0. \end{equation}
This IR condition is similar to that from top down brane models \cite{Erdmenger:2007cm} but imposed at the RG scale where the flow becomes ``on-mass-shell". Here we are treating $X(\rho)$ as a constituent quark mass at each scale $\rho$. Were we to continue the flow below this quark mass scale we would need to address the complicated issue of the decoupling of the quarks from the running function $\gamma$. 

Now we can introduce the field $Y$ that describes the operator $\bar{q} F^2 q$. It will have an intrinsic action
\begin{equation}
S   =   \int d^4x~ d \rho\,  \rho^{11} 
\left[  {1 \over r^2} |D Y|^2  
 +  {\Delta m_Y^2 \over \rho^2} |Y|^2    \right], 
\end{equation} 
where now $r^2 = \rho^2 + |X|^2 + \rho^8|Y|^{2}$. Here $Y$ has energy dimension of -3 and when $\Delta m_Y^2=0$ has the solution
\begin{equation} Y = \alpha + {\beta \over \rho^{10}} \,.\end{equation}
$\alpha$ is the source for the $\bar{q} F^2 q$ term in the action and $\beta$ has the dimension of the vev. If we include $\Delta m_Y^2$ here then the dimension of $\bar{q} F^2 q$ will run from the UV value of 7. For this toy model we will assume the dimension is $7 - \gamma_1$ so its dimension falls but the BF bound will not be violated at the scale where $\gamma_1=1$ and $X$ condenses. 

We can now include higher order terms in the action mixing the fields that favour translational symmetry breaking. For example we might include
\begin{equation} \Delta {\cal L} = \tilde{\kappa}_3 {\rho^7 \over r^2}   |f|^2  \partial_M X^\dagger \partial^M Y\,, \end{equation}
where $\tilde{\kappa}_3$ is dimensionless. 
As the vev of $f$ grows this will introduce a kinetic mixing term that will drive the lowest mass eigenstate's mass more negative by a $k$ dependent factor. As written this term tends to drive the kinetic term in the holographic $\rho$ direction negative also. However, there are terms  that break the $\rho-x$ symmetry after the $f$ field acquires a vev. For example
\begin{equation} \Delta {\cal L} = \tilde{\kappa}_4 {\rho^9 \over r^4}  (\partial_M f \partial^M X^\dagger)
(\partial_N f \partial^N Y)\,. \end{equation}
$\tilde{\kappa}_4$ is again dimensionless. Since $f$ only has a non-zero $\rho$ dependence this term is, on substituting the vev, simply a correction to the $\rho$ derivative term mixing $X$ and $Y$. By picking $\tilde{\kappa}_4$ appropriately ($\tilde{\kappa}_3 = - 16 \tilde{\kappa}_4$) one can remove the mixing term in the $\rho$ derivative but leave a mixing term in the $x,t$ coordinates
\begin{equation} \Delta {\cal L} = \tilde{\kappa}_2 {c^2 \over \rho r^4}   \partial_\mu X^\dagger \partial^\mu Y \,.\end{equation}
For our computation below we will assume that the correction to the $\rho$ kinetic term is zero and that $\tilde{\kappa}_2$ is our free parameter. 

In such a model one can numerically solve the coupled ODEs for the profiles of $X$ and $Y$ and then evaluate the action on those solutions to determine the effective potential of a solution. Performing this computation for a solution of the form $X / Y \sim f_{X/Y}(\rho) \sin k x$ allows one to plot the potential against $k$.  For example, to set the runnings we can study $N_c=3$ $N_f=3$ (of course QCD with these values does not generate stripes but these choices are indicative of the behaviour), with the scale at which $\gamma=1$ to be $\Lambda_{QCD}$  and set $c = \Lambda^4_{QCD}$. In Figure 1 we plot the potential as a function of $k^2$ for different choices of the higher dimension operator's coefficient. We see that for ${\cal O}(1)$ negative values an instability for stripes is indeed present. Strictly for QCD, which we know respects Lorentz invariance, we have placed limits on $\tilde{\kappa}_2$ by this argument. The instability mechanism may be present in other gauge theories though. 

\begin{figure}[]
\centering
\includegraphics[width=8.5cm]{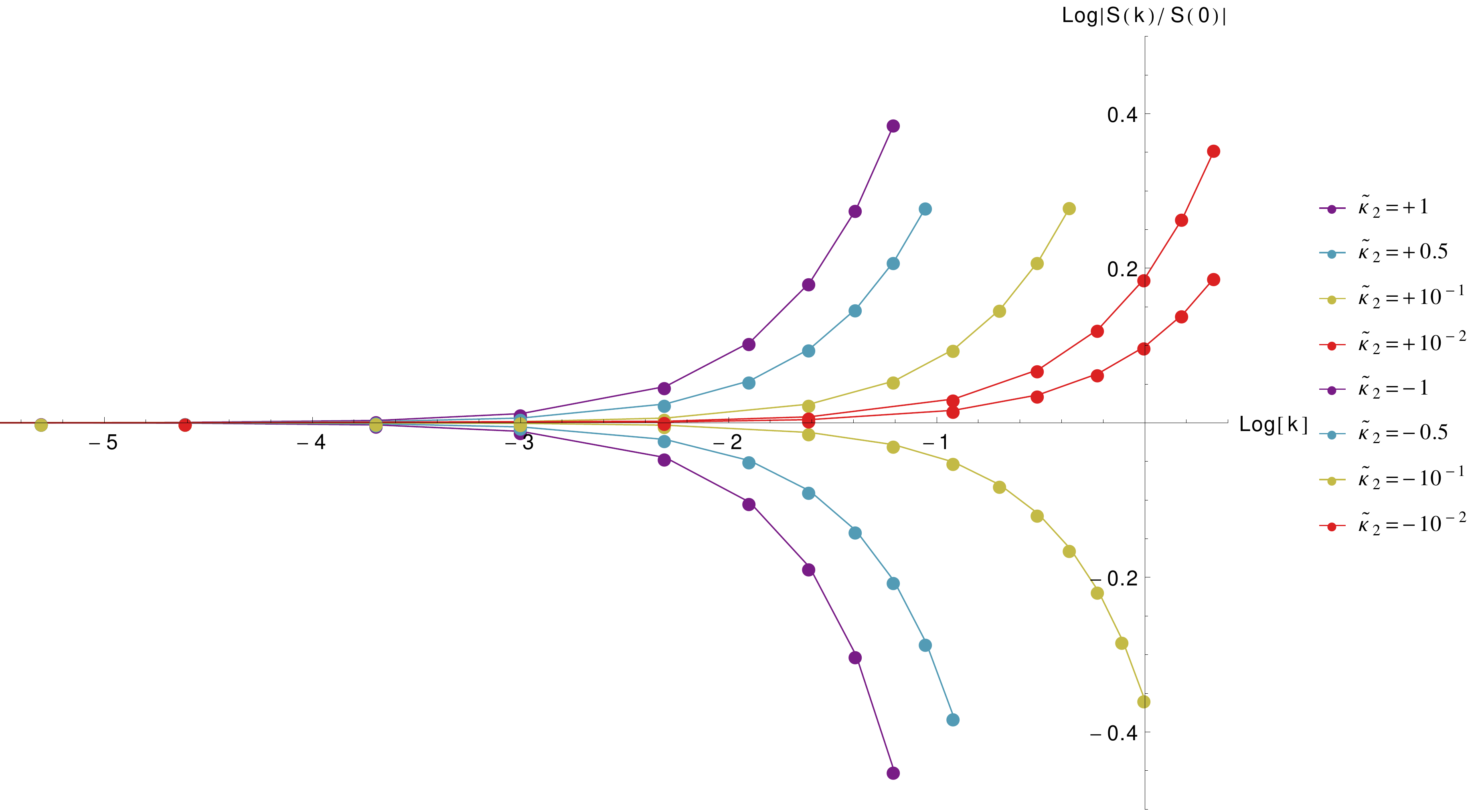}
\caption{Potential (normalized by that at $k=0$) against $\ln[k/\Lambda_{QCD}]$ for varying values of the coefficient of the higher dimension operator (which is a mix of $\tilde{\kappa}_2$ and $\tilde{\kappa}_3$. We set here $N_c=3$ $N_f=3$, the scale at which $\gamma=1$ to be $\Lambda_{QCD}$  and set $c = \Lambda^4_{QCD}$.}
\label{massfull}
\end{figure}

At this point we will cease speculating about such unknown gauge dynamics and simply assume that field theories with translational symmetry broken in the vacuum exist. We will explore whether they are phenomenologically interesting and viable as part of Beyond the Standard Model physics.

\section{Striped Phases and the  Cosmological Constant in $R^2$ Gravity}

Our interest in such striped, chequer-board-like or cuboid phases is that they could have a dramatic cosmological consequence. The basic observation is that the response of the metric to such an inhomogeneity in the mass-energy distribution will be replaced by some average effect on scales much larger than $1/\Es$, however since the dynamical equations for the metric are non-linear, this averaging does not lead to the same dynamical equations for some `average' metric but rather results in corrections to the equations themselves. 

The fact that inhomogeneity can in general result in a ``cosmological back-reaction'' has been widely investigated, see {\it e.g.} refs. \cite{Rasanen:2003fy}-\cite{Buchert:2011sx}.  These papers were inspired by the possibility that cosmological inhomogeneity  (the fact that matter is not uniformly distributed at scales smaller than about 100 Mpc, but instead concentrated in walls and clusters of galaxies, containing stars and planets {\it etc.}) leads to corrections to the average expansion rate which could explain the observations that indicate that the universe is currently undergoing accelerated expansion, such as \cite{Hinshaw:2012aka}.  (The underlying assumption is that for some still mysterious reason the fundamental cosmological constant exactly vanishes, either because the quantum field theoretic dark energy also vanishes or because somehow gravity is decoupled from it.) However to date the results of these studies have been either negative or inconclusive. 

One particularly elegant and clean approach to inhomegeneity in a cosmological context,  was put forward by Green and Wald \cite{Green:2010qy,Green:2013yua}. A brief summary of the analysis is as follows: one splits the metric as
\begin{equation} g_{ab} = g^{(0)}_{\a\b} + h_{\a\b}\,, \end{equation}
where $g^{(0)}_{\a\b}$ is the Freedman-Robertson-Walker metric of standard cosmology and $h_{\a\b}$ is the piece sensitive to the matter distribution which here we imagine is the stripey phase of the gauge theory with structure on scale $\Lambda_{stripe}$. The $R_{\a\b}-\half g_{\a\b} R$ terms in the equation of motion split into the standard ones for  $g^{(0)}_{\a\b}$ plus extra pieces dependent on $h_{\a\b}$. The philosophy is to take the spatial average of the pieces dependent on  $h_{\a\b}$ and then treat the resulting terms as an effective addition to the stress energy tensor of the matter content of the Universe. Assuming that a certain weak limit exists, they perform a rigorous diffeomorphism-invariant averaging process of the gravitational response to mass density fluctuations through the application of this weak limit. Assuming that the matter stress-energy tensor $T_{\alpha \beta}$ satisfies the weak energy condition, Green and Wald prove that 
the averaged effect of the coupled matter plus gravitational fluctuations is then encoded in  this limit in an additive correction $\tz_{\alpha \beta}$
to the stress-energy which is traceless and also satisfies the weak energy condition. They therefore identify it 
with gravitational radiation.
In particular 
in a FLRW background metric, $\tz_{\alpha\beta}$ is diagonal, corresponding to an effective fluid with pressure $p=\rho/3\ge0$, leading to the conclusion that such a back-reaction cannot mimic dark energy. 

The situation changes dramatically however if we now entertain the possibility that Einstein's General Relativity equations themselves already have gravitational corrections. Indeed, now that the (galactic foreground) dust has settled on the BICEP2 debate, it is clear that the Starobinsky model of ``$R^2$ inflation'' \cite{Starobinsky:1980te,Davies:1977ze}, one of the earliest models of inflation, remains highly favoured observationally. In this model the Lagrangian density is given by
\beq \call =
\frac{1}{2\kappa}
 \left(R+ 
 \frac{R^2}{6M^2} 
 \right)+\mathcal{L}_{Matter}
\eeq
(where $\kappa=8\pi G$, and $\sqrt{-g}$ is included in the measure). The new parameter is the so-called scalaron mass, which must be $M\approx 3\times10^{13}$ GeV, in order to agree with cosmological observations. Following the above philosophy we continue to assume that the underlying (quantum field theoretic) net vacuum energy effectively vanishes (hence the absence of a cosmological constant term above). 
Back-reaction is again encoded in a diffeomorphism invariant effective additive correction $\tz_{\alpha\beta}$ to the matter stress energy tensor, however it is now  not traceless. Instead \cite{Preston:2014tua}
\beq
\label{tz}
\kappa\, t^{(0)} \underset{weak}{=} -\frac{{R^{(1)}}^2}{6M^2}\,,
\eeq
where $R^{(1)}$ is the linearised Ricci scalar of the gravitational fluctuations $h_{\a\b}$, and equality holds rigorously in a certain weak limit. Encouragingly, $\tz$ thus must be negative, in agreement with the current acceleration of the universe. 

In general the requirement of a weak limit combined with subtle cancellations up to second order in the fluctuations, make it difficult to compute $\tz_{\mu\nu}$ in any given model \cite{Green:2013yua}. This is illustrated by the existence of an equally valid alternative expression \cite{Preston:2014tua}:
\beq
\kappa\,\tz  \underset{weak}{=} \frac14hR^{(1)}-\frac12h^{\alpha\beta}R^{(1)}_{\alpha\beta}\,.
\eeq
However in \eqref{tz}, cancellations have been manifestly achieved in as much as it leaves behind a negative definite term. Furthermore it behaves parametrically in the right way, in the sense that if we assume that the fluctuations are generated independently of the scalaron scale, then we recover the tracelessness of the additive correction \cite{Green:2010qy,Green:2013yua} in the limit $M\to\infty$. Therefore it seems reasonable to apply order of magnitude estimates to \eqref{tz} in order to obtain a rough estimate of the effective cosmological constant generated by gravitational back-reaction from a striped phase. Setting $R^{(1)}\sim \partial^2 h \sim \kappa \rho$ where $\rho$ is the local mass over-density, we have from \eqref{tz} that the effective vacuum energy $\tz \sim -\kappa \langle\rho^2\rangle/(6 M^2)$. Now we recognise that $\kappa=1/\MP^2$, where the reduced Planck mass is $\MP=2.44\times10^{18}$ GeV, that $\tz=-\EV^4$ where the current effective vacuum energy is $\EV= 10^{-12}$ GeV in order to agree with observations, and finally that the RMS value  $\sqrt{\langle \rho^2\rangle}\sim \Es^4$, where $\Es$ is the energy scale that sets both the amplitude and wavelength of the striped phase. Combining, we therefore find that
\beq
\EV\sim \frac{\Es^2}{\sqrt{6M\MP}}\,,
\eeq
from which we deduce that $\Es\sim  140$ GeV, intriguingly close to the Higgs' mass and the EW (electroweak) scale. Of course if $\tz$ is to mimic a cosmological constant and thus drive the present day acceleration of the universe, it must also be (at least approximately) constant. However this is guaranteed by the present mechanism since as the universe expands the stripes are not diluted  but instead rearrange and get created to fill `the gaps', since the wavelength is set at $\Es$ by the microscopic dynamics described in the previous section.

We note again that we have \underline{not} suggested a mechanism that naturally suppresses large contributions to the cosmological constant, in other words we are not attempting to solve the infamous cosmological constant problem. Nevertheless we have still shown how to generate a new type of contribution that can be significant, indeed can be sufficiently large to explain on its own the value deduced from the present day cosmological acceleration.

\section{Constraints on Lorentz Symmetry Breaking}

We have seen that a striped phase of a gauge theory vacuum could potentially be responsible for generating the observed cosmological constant and that in Starobinsky inflation models a characteristic scale close to the EW scale is implied. In this section though we will address the stringent phenomenological constraints on Lorentz violation and show that this is catagorically ruled out outside of a totally disconnected dark sector of the universe.

Let us first entertain the idea that the Higgs vev or some component of it is spatially varying. Consider an induced  mass term for some scalar that varies spatially: 
\[
m^2(x)= m_0^2 + \mu^2(x) \,,
\]
where for simplicity we take the variation to occur along a fixed axis defined by a fixed (space-like) four-vector $k^\mu$: $\mu^2(x) = \mu_0^2 \cos(k^\mu x_\mu)$ and the magnitude of the wave number $|k|\sim \Es$. Write the propagator $(\Box + m_0^2)^{-1}$ in position space as $G(x_1,x_2)$.
By expanding the full inverse propagator
\bea
&&(\Box+m^2)^{-1}(x_1,x_2) = G(x_1,x_2) \nonumber\\
&&- \int_{y_1} G(x_1,y_1) \mu^2(y_1) G(y_1,x_2)  \nonumber\\
&&+\int_{y_1,y_2} G(x_1,y_1) \mu^2(y_1) G(y_1,y_2)\mu^2(y_2)G(y_2,x_2)\nonumber\\
&&+\cdots\,,
\eea
and recognising that $\mu^2(y_n)= \frac{\mu_0^2}{2}({\rm e}^{ik\cdot y_n}+{\rm e}^{-ik\cdot y_n})$ is a standing superposition of plane waves, we see that the modulated mass-term is physically equivalent to a sum of tree level processes that exchange momentum $\pm k^\mu$ with the (spatially varying) vacuum.

The most dramatic effect of the spatially varying terms is to lead to an apparent break-down of momentum conservation  along the $k^\mu$ axis such that at each successive occurrence of $\mu^2(y_n)$ the four-momentum  receives a kick  $\Delta p^\mu = \pm k^\mu$. (Similar to Umklapp scattering in a crystal, the momentum is actually transferred to the condensate.) The effect of this kick is generically to push the particle off shell and thus the process needs completing with further interactions (\eg\ using Standard Model interactions) into on-shell decay products whose total four-momentum equals the shifted value. Assuming that the particle is kicked far off shell it is straightforward to see that the amplitude for the process is $\sim \mu_0^2/k^2$.

In the diagrammatic expansion there are exchanges with the vacuum that result in a net nil transfer of momentum, for example we can first accept $+k^\mu$ from the vacuum, propagate from $y_1$ to $y_2$ and then accept $-k^\mu$ from the vacuum, and {\it vice versa}. Summing all such pair-wise processes results in a self-energy correction for the particle $m_0^2 \mapsto m^2_0+\Sigma$ where
\[
\Sigma = \frac{\mu_0^4}4\sum_\pm \frac1{(p\pm k)^2-m^2_0}\,.
\]
There are infinitely many other processes with a more complicated pattern of exchanges of $\pm k^\mu$ which are equally important unless there is a hierarchy in the parameters.
If for the moment we assume that $\mu_0^2\ll -k^2$ (recall that $-k^2>0$ since the wave-vector defines a spatial modulation), then the process we have considered is the leading process in a perturbative expansion of $\mu_0^2/k^2$. Expanding the above expression we see that for small $\mu_0^2/k^2$ the leading effects are a shift of the mass $\mu_0^4/(2k^2)$ which is probably pretty harmless and a rotational and Lorentz symmetry breaking term 
\[
\Delta \Sigma = -2 \left(\frac{\mu_0^2}{k^2}\right)^2 \frac{(p\cdot k)^2}{k^2}\,.
\]
Unfortunately such terms are severely constrained and in practice we furthermore expect that $\mu_0\sim |k| \sim \Es$ \ie\ that these terms appear with ${\cal O}(1)$ coefficients.

The strongest constraints on Lorentz violation arise from low energy Michelson-Morley style tests of the electron QED system \cite{Pospelov:2004fj,Pruttivarasin:2014pja}. The leading Lorentz violating term is
\begin{equation}
{\cal L} = {1 \over 2} i \bar{\psi} (\gamma_\nu + c_{\mu \nu} \gamma^\mu) D^\nu \psi - \bar{\psi} m_e \psi\,,
\end{equation}
where $c_{\mu \nu}$ is a symmetric fixed tensor in the laboratory frame describing the Lorentz violation. The components of $c_{\mu \nu}$ are constrained at the level of $10^{-18}$. Generically one would expect the electron-photon vertex to be corrected by W and Z exchange which in turn would link to the Higgs vev and generate such terms if the EW scale did generate striped behaviour. The maximum suppression is only of order weak scale couplings to the fourth power (assuming the vev of the Higgs and the scale of the stripes are close as is natural) and such terms are nothing like sufficiently suppressed. 

{To escape tight contraints such as these one can place the Lorentz symmetry breaking into a dark sector. Standard Model particles at a scale $q^2$ see effects from Planck suppressed operators suppressed by a factor of $q^2/M_{Pl}^2$. However, here the Planck scale physics respects Lorentz invariance which is only broken in the dark sector by the striped gauge dynamics. The gravitational sector must therefore communicate to that new sector and an additional suppression of $v^2/M_{PL}^2$ enters. For scales $v$ as low as the weak scale in the dark sector this factor becomes a suppression by $32$  further orders of magnitude. The effect can certainly be made invisible, although of course this is a disappointing way to proceeed since one would hope for a signal.

Finally we again stress that striped or chequer-board configurations will have Goldstone states associated with the spontaneous breaking of Lorentz symmetry that propagate freely in the unbroken directions \cite{Low:2001bw}. Such states clearly do not exist in the visible sector but potentially could in a dark sector. For a cuboid  configuration the mass term vanishes only at the corners of the cubes 
and thus the potential Goldstone mode cannot propagate. Similar comments apply to more general crystal-like phases.

In conclusion then, if translational symmetry breaking is responsible for the observed cosmological constant it must necessarily be present only in a sector of physics disconnected from the Standard Model. Of course in principle new dark sectors could have been cast off  at scales from the Standard Model sector up to the Planck scale so sufficient suppression could be achieved but the hope of an experimental signature are then damped and the intriguing link to EW physics through generation of an effective cosmological constant would also be lost.

\bigskip

\section{Discussion}

Recent work has shown that short scale structure can manifest in $R^2$ gravity as a cosmological constant on long distance scales \cite{Preston:2014tua}. To produce such structure that does not dilute with the expansion of the Universe one should attempt to embed the structure in the vacuum. This motivated us to consider how Lorentz violating striped phases could appear in field theory. We have argued that higher dimension operators in theories with condensation near the theories' UV cutoff could cause the spontaneous breaking of Lorentz symmetry. This naturally maps to the low energy description of asymptotically free gauge theories where many operators are involved in the vacuum structure and where higher dimension operators are naturally produced by the strong coupling scale. We displayed a holographic model of such a scenario showing the instability to Lorentz violating phases. These ideas raise the possibility that exotic gauge theories might exist that break Lorentz invariance spontaneously and it would be interesting to explore this in lattice simulations of models beyond the Standard Model.

In Starobinsky inflation models with a vanishing fundamental cosmological constant, the scale of stripes needed to generate the observed current cosmological constant is of order the electroweak scale. However, as we have discussed, low energy constraints on Lorentz violation in the electron photon system  do not allow any such Lorentz violation in the visible sector. A deeply disconnected dark sector could still display the phenomenon. These arguments provide further motivation to continue experimental tests of Lorentz symmetry. 

\bigskip

\noindent {\bf Acknowledgements:} The authors are grateful for the support of an STFC consolidated grant.

\end{document}